\documentclass[%
 reprint,
showpacs,preprintnumbers,
bibnotes,
 amsmath,amssymb,
 aps,
]{revtex4-1}

\usepackage{graphicx}% Include figure files
\usepackage{dcolumn}% Align table columns on decimal point
\usepackage{bm}% bold math
\usepackage{multirow}
\usepackage{caption}
\usepackage{subcaption}
\usepackage{tikz}
\usepackage{subfig}
\usepackage{subfloat}
\usepackage{amsmath}

\begin{document}
%\preprint{APS/123-QED}

\title{Polymer as a function of monomer: Analytical quantum modeling}% Force line breaks with \\

\author{Mohammad \surname{Nakhaee}}
%can be forced with \\
 \email{m.nakhaee@std.du.ac.ir}
\author{S. Ahmad \surname{Ketabi}}
%can be forced with \\
\affiliation{Damghan University, Damghan, Iran}%Lines break automatically or 
%can be forced with \\



%\author{}
%\affiliation{
%\affiliation{
%Third institution, the second for Charlie Author
%\affiliation{%


\date{\today}% It is always \today, today,

\begin{abstract}
To identify an analytical relation between the properties of polymers and their's monomer a Metal-Molecule-Metal (MMM) junction has been presented as an interesting and widely used object of research in which the molecule is a polymer which is able to conduct charge. The method used in this study is based on the Green's function approach in the tight-binding approximation using basic properties of matrices. For a polymer base MMM system, transmission, density of states (DOS) and local density of states (LDOS) have been calculated as a function of the hamiltonian of the monomer. After that, we have obtained a frequency for LDOS variations in pass from a subunit to the next one which is a function of energy.
\end{abstract}

\pacs{73.23.Ad, 73.63.Rt, 73.22.-f, 75.75.-c, 85.75.-d}% PACS, the Physics and Astronomy
\keywords{Molecular junction, Polymer, Monomer, Tight-binding, Green's function}%Use showkeys class option if keyword
\maketitle

%\tableofcontents

\section{Introduction}
\label{Introduction}
\begin{figure}[!ht]
\centering
\includegraphics[scale=1]{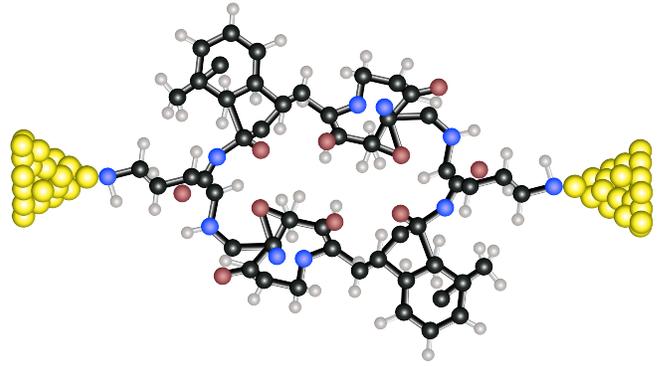}
\caption{A schematic figure of a metal-molecule-metal system.} 
\label{mmm}
\end{figure}

Molecular electronics have been developed to realize biodegradable and large-area electronic products. These devices have been fabricated with various usges. As an instance, J.E. Lilienfeld invented a method which was proposed to control electric currents.\cite{Lilienfeld} This method led to introduction of field-effect transistor which operates as a capacitor between a source and a drain electrode. In 1987, the first organic field-effect transistor was made by Koezuka and co-workers based on a polymer that is able to conduct charge, eliminating the need to use expensive metal oxide semiconductors.\cite{Koezuka} Polymers have recently gained great interest as building blocks for electronic applications that require low-cost, and potentially flexible form–factors. Usually, polymer synthesis methods are extremely affordable and apart from the technical physical considerations. There are many works on polymers and their applications such as drivers for electronic papers,\cite{Gelinck} radio-frequency recognition,\cite{Subramanian} and circuits for flat displays.\cite{Zhou} When processed into thin films, polymers can be used to build transparent materials, taking advantage of their weak absorbance in the visible spectrum. Invisible electronic circuits based on polymers are important for the uses requiring transparency, such as sensors and bendable displays.~\cite{Nomura2003,Nomura2004,Forrest,Reuss,Wager}\\
Recently, Metal-Molecule-Metal (MMM) structures (Figure \ref{mmm}) have attracted scientists. Their broad range of applications have led chemists, physicists, and other scientists and engineers engaged in organic electronics to produce devices with new abilities and have improved efficiencies relative to their primary counterparts~\cite{Nitzan,Liang,MMM1,MMM2,MMM3}. Nowadays, using polymers in electrical devices is a point of strength. Specifically, we are interested in discussing around a MMM junction where the polymers serves as the center part. \\
Molecular properties include any of a material's properties that becomes evident during, or after, a chemical reaction and any property that is measurable, whose value describes a state of a physical system which are supervenient on the underlying atomic structure, which may in turn be supervenient on an underlying quantum structure. It may be difficult to determine whether a given property which can be seen and measured is a property that is emerged from quantum confinement or not. The best way to understand it is to get stuck into the theories. Using this method of understanding can link properties together, even if one doesn't have enough information about the story and the fact which is happened within the system.\\
The rest challenge now is to find out how to express different properties of a polymer in terms of the ones of it's monomer. These properties are supervenient on the underlying atomic structure and the quantum indicators of the monomer. Therefore, physical properties of polymers can be categorized as being either dependent or independent of the size of the polymer.\\
\begin{figure*}[!ht]
\centering
\includegraphics[scale=0.54]{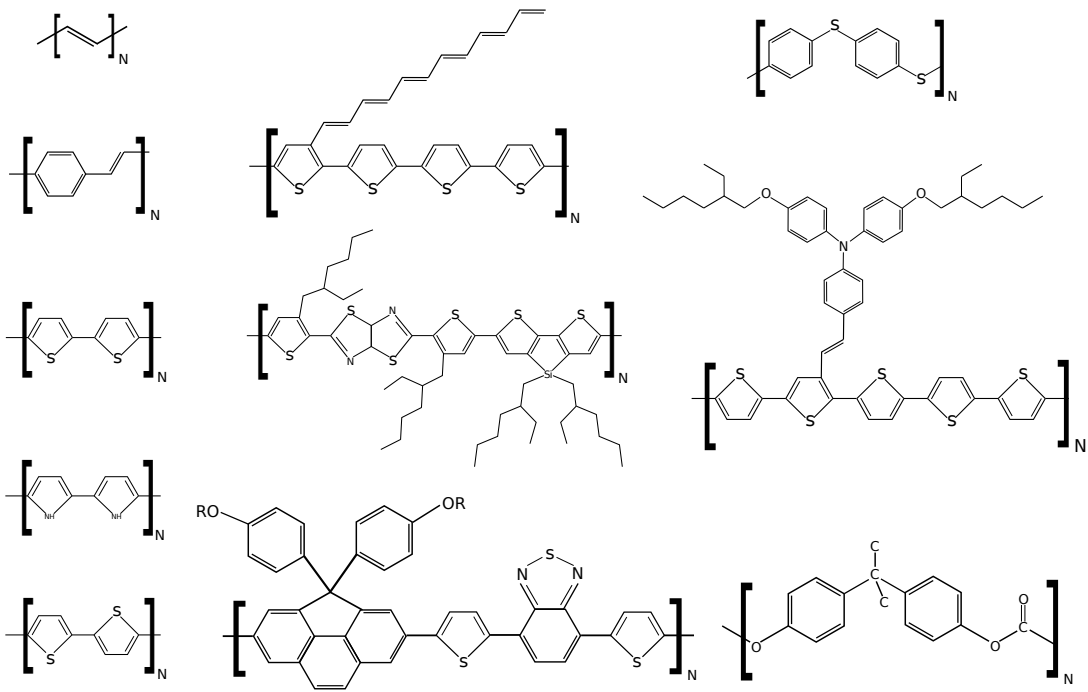}
\caption{Polymeres which have one bond between repeated subunit.}
\label{polymers}
\end{figure*}
Due to the wideness of the types of polymers, the existence of theories and computational methods will be a big help for scientific progress. Figure (\ref{polymers}) shows some polymers which are defined by a repeated subunit. We are interested in the properties which are widely studied in many ways. The tight-binding or linear combination of the atomic orbitals~\cite{Andersen,Ackland,Goringedag} is one of the most important methods which is a semi-empirical method that is primarily used to calculate the eigen energies of the system. This method is simple and computationally very fast.\\
The purpose of this essay is to introduce a new method for quantum simulation of the polymers as a function of their monomer. For the moment, this general outline may be sufficient, and the main approach and important topics, including the newly calculated properties, would be explained in the rest of the paper. In each section, we report on the present status of the approaches used in detail.
%%%%%%%%%%%%%%%%%%%%%%%%%%%%%%%%

\section{Method}
\label{method}
In this section we are going to introduce an analytical method based on basic properties of matrices to calculate the properties of a polymer base system via Green's theorem. The new method gives us several benefits with realy less computational effort than other methods, which could facilitate significantly the computations. Because of the wide variety of polymers, it is helpful to classify them in a way that polymers are common on the number of the connections between repeated subunit. As shown in the figure (\ref{polymers}) many of polymers have one bond between monomers. This subtle physical note is the basis point of this method. The most commonly used device geometry is a MMM system. Green's function theory is selected to establish this method as a powerful and flexible approach for applying it to many types of polymers as the center part of the MMM system. To achieve this purpose we would write the general hamiltonian in tight-binding approximation, for such a polymer which contains $ N $ monomers, via second quantization representation as follows
\begin{small}
\begin{widetext}
\begin{align}\label{EqHamiltonian}
\mathcal{H} = \sum_{m,i} \epsilon_i c_{i+(m-1)n}^\dagger c_{i+(m-1)n} + \sum_{m,i,j}^{'} t_{i,j} c_{i+(m-1)n}^\dagger c_{j+(m-1)n} + \sum_m t_l(c_{m n+1}^\dagger c_{m n}+c_{m n}^\dagger c_{m n+1})
\end{align}
\end{widetext}
\end{small}
where the summations run over all atoms of the polymer. The prime on summation symbol omits the cases $i=j$. The parameter $\epsilon_i$ is the energy of the electrons at site $i$, the complex numbers $t_{i,j}=t_{j,i}^{*}$ is the transfer energy between site $i$ and $j$, $c^\dagger_i$ ($c_j$) is the creation (annihilation) operator of electrons at site $i$ ($j$), $ m $ stands on the number of the subunit ($ m=[1:N]$) and $ t_l $ (refers to the link between monomers) is the transfer energy between the last atom of a subunit and the first atom of the next subunit. The equation \ref{EqHamiltonian} can be written in matrix form as follows

\begin{small}
\begin{equation}\label{MainMatrix}
\mathcal{H}=\left(
\begin{array}{ccccccc}
 H & T & \mathit{0} & \cdots & \mathit{0} & \mathit{0} & \mathit{0} \\
 T^{\dagger} & H & T & \cdots & \mathit{0} & \mathit{0} & \mathit{0} \\
 \mathit{0} & T^{\dagger} & H & \ddots & \mathit{0} & \mathit{0} & \mathit{0} \\
 \vdots & \vdots & \ddots & \ddots & \ddots & \vdots & \vdots \\
 \mathit{0} & \mathit{0} & \mathit{0} & \ddots & H & T & \mathit{0} \\
 \mathit{0} & \mathit{0} & \mathit{0} & \cdots & T^{\dagger} & H & T \\
 \mathit{0} & \mathit{0} & \mathit{0} & \cdots & \mathit{0} & T^{\dagger} & H \\
\end{array}
\right)_{N\times N}
\end{equation}
\end{small}
Let us to introduce some compact notations to explain a matrix. As usual $ A_{i,j} $ is the element of the matrix $ A $. Colon symbol is applied to extract parts of a matrix. Generally, $ A_{i:j,m:n} $ is the submatrix formed from rows $ i $ through $ j $ and columns $ m $ through $ n $ of the matrix $ A $, the matrix $ A_{i:j} $ is equal to the square matrix $ A_{i:j,i:j} $ and $ A_{i} $ is a more compact form of $ A_{i:n_A} $ in which $ n_A $ is the dimention of the matrix $ A $. The order of the full matrix in the equation \ref{MainMatrix} is $ n N \times n N $ in which $ n $ is the number of atoms of the monomer. The matrix $ H $ is the hamiltonian of the monomer and it's order is $ n \times n$. Recalling once again that, generally, the non-vanish element of the matrix $ T $ is complex and defined by the equation

\begin{small}
\begin{equation}\label{Tlink}
T_{i,j}=t_l\delta_{i,n}\delta_{j,1}
\end{equation}
\end{small}

The idea is to calculate the determinant of the matrix $ \mathcal{H} $ and other properties using the determinant. Now we define the new matrix $ \mathcal{M}^i $ (the raising index is used to not have conflict with definition of the lowering index) as follows
\begin{small}
\begin{equation}\label{MatrixM}
\mathcal{M}^i=\left(
\begin{array}{cc}
 \mathit{1} & R_{1 \times (n N-i+1)} \\
 C_{(n N-i+1) \times 1} & \mathcal{H}_{i}\\
\end{array}
\right)
\end{equation}
\end{small}

In which the elements of the matrices $ R $ and $ C $ are defined as
\begin{small}
\begin{align}\label{def1}
R_{1,i}&=t_l\delta _{n-k+1,i} \nonumber \\
C_{i,1}&=\delta _{i,n-k+2}
\end{align}
\end{small}

The matrix $ \mathcal{M}^i $ is a dummy matrix which helps us to achieve our purpose. The determinant of the matrix $ \mathcal{M}^{k n-j+1} $ can be calculated in two ways as follows
\begin{small}
\begin{align}\label{dets}
\left| \mathcal{M}^{k n-j+1}\right| &=\left| H_{j}\right| .\left| \mathcal{H}_{k n+1}\right| \nonumber \\
\left| \mathcal{M}^{k n-j+1}\right| &=\left| \mathcal{H}_{(k-1)n+j}\right| +t_l{}^2\left|H_{j:n-1}\right| .\left| \mathcal{H}_{k n+2}\right|
\end{align}
\end{small}

We can eliminate the variable $ \left| \mathcal{M}^{k n-j+1}\right| $ by subtraction of the two equations. In other words, we can get a recursive equation as following
\begin{small}
\begin{align}\label{RecursiveEq}
\left| \mathcal{H}_{(k-1)n+j}\right| =\left| H_{j}\right| .\left| \mathcal{H}_{k n+1}\right| -t_l{}^2\left|H_{j:n-1}\right| .\left| \mathcal{H}_{k n+2}\right|
\end{align}
\end{small}

In which the domain of $ k $ is $ [1:N] $, the range of $ j $ is $ [1:n] $, and therefore the index $ (k-1)n+j $ runs through $ [1:n N] $. One can expand the equation \ref{RecursiveEq} (by the parameter $ j $) to a set of $ n $ linear equations in $ n $ unknowns determinants. These equations are recursive and can be put in the form of a matrix-vector equation by iterate this procedure $ N-1 $ times for the parameter $ k $ as follows

\begin{scriptsize}
\begin{align}\label{MatrixEq}
\left(
\begin{array}{c}
 \left| \mathcal{H}_1\right|  \\
 \left| \mathcal{H}_{2}\right|  \\
 \vdots \\
 \left| \mathcal{H}_{n-1}\right|  \\
 \left| \mathcal{H}_{n}\right|  \\
\end{array}
\right)=U^{N-1}\left(
\begin{array}{c}
 \left| \mathcal{H}_{n N-n+1}\right|  \\
 \left| \mathcal{H}_{n N-n+2}\right|  \\
 \vdots \\
 \left| \mathcal{H}_{n N-1}\right|  \\
 \left| \mathcal{H}_{n N}\right|  \\
\end{array}
\right)=U^{N-1}\left(
\begin{array}{c}
 \left| H_{1}\right|  \\
 \left| H_{2}\right|  \\
 \vdots \\
 \left| H_{n-1}\right|  \\
 \left| H_{n}\right|  \\
\end{array}
\right)
\end{align}
\end{scriptsize}

In which $ U $ is defined as following matrix

\begin{scriptsize}
\begin{align}
U=\left(
\begin{array}{ccccccc}
 \left| H_{1}\right|  & -t_l{}^2\left| H_{1:n-1}\right|  & 0 & 0 & \cdots & 0 & 0 \\
 \left| H_{2}\right|  & -t_l{}^2\left| H_{2:n-1}\right|  & 0 & 0 & \cdots & 0 & 0 \\
 \vdots & \vdots & \vdots & \vdots & \cdots & \vdots & \vdots \\
 \left| H_{n-1}\right|  & -t_l{}^2\left| H_{n-1:n-1}\right|  & 0 & 0 & \cdots & 0 & 0 \\
 \left| H_{n}\right|  & -t_l{}^2 & 0 & 0 & \cdots & 0 & 0 \\
\end{array}
\right)_{n\times n}
\end{align}
\end{scriptsize}

Now, the challenge is to raise the square matrix $ U $ in the equation \ref{MatrixEq} to the power of $ (N-1) $. One can change the basis of the matrix representation for which the matrix takes the diagonal form and after raising the matrix to the power of $ (N-1) $, change the basis of the new matrix to what it was before. So, the equation \ref{MatrixEq} becomes as follows

\begin{tiny}
\begin{align}\label{MatrixEq2}
\left(
\begin{array}{c}
 \left| \mathcal{H}_1\right|  \\
 \left| \mathcal{H}_{2}\right|  \\
 \vdots \\
 \left| \mathcal{H}_{n-1}\right|  \\
 \left| \mathcal{H}_{n}\right|  \\
\end{array}
\right)=X.\left(
\begin{array}{ccccc}
 \lambda_1^{N-1}  & 0  & 0 & \cdots & 0\\
 0  & \lambda_2^{N-1}  & 0 & \cdots & 0\\
 0  & 0  & 0 & \cdots & 0\\
 \vdots & \vdots & \vdots & \ddots & \vdots \\
 0  & 0  & 0 & \cdots & 0\\
\end{array}
\right).X^{-1}.\left(
\begin{array}{c}
 \left| H_{1}\right|  \\
 \left| H_{2}\right|  \\
 \vdots \\
 \left| H_{n-1}\right|  \\
 \left| H_{n}\right|  \\
\end{array}
\right)
\end{align}
\end{tiny}

In which $ \lambda_1 $ and $ \lambda_2 $ are non-zero eigenvalues of the matrix $ U $, and $ X $ is a matrix of the set of eigenvectors. The variables $ \lambda_1 $ and $ \lambda_2 $ would be obtained as following equations

\begin{small}
\begin{align}\label{eigenvalues}
\lambda _1=\frac{1}{2} \left(\left| H\right| -t_l{}^2\left| H_{2:n-1}\right| -\Delta
\right)\\
\lambda _2=\frac{1}{2} \left(\left| H\right| -t_l{}^2\left| H_{2:n-1}\right| +\Delta \right)
\end{align}
\end{small}

In which $ \Delta $ can be calculated as

\begin{small}
\begin{align}\label{Delta}
\Delta =\sqrt{\left(\left| H\right| +t_l{}^2\left| H_{2:n-1}\right| \right){}^2-4t_l{}^2\left|
H_{1:n-1}\right|  \left| H_{2:n}\right|  }
\end{align}
\end{small}

Also, $ X $ should be obtained as following matrix

\begin{small}
\begin{align}\label{XMatrix}
X=\left(
\begin{array}{ccccc}
 u_1(\lambda_1,\lambda_2) & v_1(\lambda_1,\lambda_2) & 0 & \cdots & 0 \\
 u_2(\lambda_1,\lambda_2) & v_2(\lambda_1,\lambda_2) & 0 & \cdots & 0 \\
 u_3(\lambda_1,\lambda_2) & v_3(\lambda_1,\lambda_2) & 1 & \cdots & 0 \\
 \vdots & \vdots & \vdots & \ddots & \vdots \\
 u_n(\lambda_1,\lambda_2) & v_n(\lambda_1,\lambda_2) & 0 & \cdots & 1 \\
\end{array}
\right)
\end{align}
\end{small}

In which $ u_i $ and $ v_i $ would be defined as follows

\begin{small}
\begin{widetext}
\begin{align}\label{eigenvectors}
u_i(\lambda_1,\lambda_2) = v_i(\lambda_2,\lambda_1) = \left| H_{i}\right| \left(\left| H\right| +t_l^2\left| H_{2:n-1}\right|
+\lambda _1-\lambda _2\right)- 2t_l^2 \left| H_{2}\right| .\left| H_{i:n-1}\right| /\left(\left| H_{n}\right|
\left(\left| H\right| +t_l^2+\lambda _1-\lambda _2\right)-2t_l^2 \left| H_{2}\right| \right)
\end{align}
\end{widetext}
\end{small}

Now, we define two new functions \begin{small}$ \widetilde{\mathcal{H}} = \widetilde{\mathcal{H}}(\epsilon) = \epsilon - \mathcal{H} - \hat{\Sigma}_l $\end{small} and \begin{small}$ \widetilde{H} = \widetilde{H}(\epsilon) = \epsilon - H(\epsilon) - \hat{\Sigma}_r $\end{small} which are altered by the interactions of metals in the MMM system. To switch the open system to a MMM system, non-interacting hamiltonians must be replaced with the new ones. The idea is to add the interaction of the left (right) lead as a self-energy $ \Sigma_l $ ($ \Sigma_r $) to the on-site energy of the first (last) atom of the polymer. It should be noted that, the hamiltonian $ H $ must be replaced by $ \widetilde{H} $ only in the last vector of the right hand side of the equation \ref{MatrixEq2}. One can find the relation between the determinants of the new hamiltonians and the old ones as follows

\begin{small}
\begin{subequations}
\begin{align}
\left| \widetilde{H}\right| &= \left| H\right| + \Sigma_r \left| H_{1:n-1}\right| \label{selfenergies1} \\
\left| \widetilde{\mathcal{H}}\right| &= \left| \mathcal{H}\right| + \Sigma_l \left| \mathcal{H}_{2}\right| \label{selfenergies2}
\end{align}
\end{subequations}
\end{small}

One can substitute the expression \ref{selfenergies1} into the equation \ref{MatrixEq2} and multiply matrices to get $ \mathcal{H} $ and $ \mathcal{H}_2 $. Using the equation \ref{selfenergies2}, the expression for $ \widetilde{\mathcal{H}} $, after some simplifications, is given by the following

\begin{small}
\begin{align}\label{detH}
\left| \widetilde{\mathcal{H}}\right| =S_{\lambda _1\lambda _2}+S_{\lambda _2\lambda _1}
\end{align}
\end{small}

In which $ S_{\lambda _1\lambda _2} $ is defined as follows

\begin{small}
\begin{widetext}
\begin{align}
S_{\lambda _1\lambda _2}&=\frac{\lambda _1{}^{N-1}}{4 \left(\lambda _1-\lambda _2\right) t_l^2 \left| H_{2}\right| } \biggl[ \left| H\right| {}^2 + \left| H_{2}\right|  \left(\Sigma _l\left(\lambda _1-\lambda _2\right)-t_l^2\left(\Sigma _l \left| H_{2:n-1}\right| +2 \left| H_{1:n-1}\right| \right)\right) \nonumber \\
&+ \left| H\right|  \left(\Sigma _l \left| H_{2}\right| +t_l^2 \left| H_{2:n-1}\right| + \lambda _1-\lambda _2\right) \biggr] \biggl[ \Sigma _r \left(t_l^2\left| H_{2:n-1}\right| +\left|
H\right| +\lambda _2-\lambda _1\right)+2t_l^2  \left| H_{2}\right| \biggr]
\end{align}
\end{widetext}
\end{small}

The equation \ref{detH} gives us the determinant of the hamiltonian of a polymer which is affected by two leads. This expression is a function of some information which are lie at the homiltonian of the monomer and the hopping between the monomers. One needs the determinant $ \left| \widetilde{\mathcal{H}}\right| $ to find the inverse of the matrix $ \widetilde{\mathcal{H}} $ which is needed in order to calculate other electrical properties of the MMM system.
%%%%%%%%%%%%%%%%%%%%%%%%%%%%%%%%

\subsection{\label{Transmission} Transmission}
Green's function method is a powerful method to study the electronic properties of the MMM system. In this method, the most important property of physical observables in transport, such as electrical current, is the transmission. The tight-binding description of transmission through a junction has been interesting for scientists to observe how much an electron wave packet would transmit in a specific energy. One needs the hamiltonian of the system (equation \ref{EqHamiltonian}) which is expressed in the tight-binding approximation. The routine method based upon a formulation for the conductance in terms of Green’s functions $ G $ is the well known Fisher-Lee linear-response \cite{FisherLee} for the conductance of a finite lattice embedded between the leads. At this point it is necessary to discuss that, for the MMM system which is studied in the current work, the transmission $ T(\epsilon) $ is a function of $ G_{1,n N} $ as seen in following equation

\begin{small}
\begin{align}\label{fisherleeeq}
T(\epsilon) = 4 \Im(\Sigma_l) \Im(\Sigma_r) \left| G_{1,n N}(\epsilon)\right| {}^2
\end{align}
\end{small}

To find $ G_{1,n N} $, one can calculate the adjugate of the matrix $ \widetilde{\mathcal{H}} $ for the element $ (1,n N) $ only, and apply the following equation

\begin{small}
\begin{align}\label{inverse}
G(\epsilon) = \frac{1}{\left| \widetilde{\mathcal{H}}\right|}adj(\widetilde{\mathcal{H}})
\end{align}
\end{small}

We hence conclude that

\begin{small}
\begin{align}
G_{1,n N}=(t_l \left| H_{2:n,1:n-1}\right|)^{N-1}G^0
\end{align}
\end{small}

In which $ G^0=G^0_{1,n} $ refers to a MMM system in which the monomer is conected to the leads. It can be calculated as follows

\begin{small}
\begin{align}
G^0=-\frac{\left| H_{2:n,1:n-1}\right|}{\left| \widetilde{\mathcal{H}} \right|}
\end{align}
\end{small}

Also, the transmission function should be obtained as

\begin{small}
\begin{align}
T(\epsilon) = (t_l^* t_l \left| H_{2:n,1:n-1}\right|{}^*\left| H_{2:n,1:n-1}\right|)^{N-1} T^0(\epsilon)
\end{align}
\end{small}

and from the equation \ref{fisherleeeq} we have the following expression for the transmission of a monomer $ T^0 $.

\begin{small}
\begin{align}
T^0(\epsilon) = 4 \Im(\Sigma_l) \Im(\Sigma_r) \frac{\left| H_{2:n,1:n-1}\right|{}^* \left| H_{2:n,1:n-1}\right|}{\left| \widetilde{\mathcal{H}} \right|{}^2}
\end{align}
\end{small}

%%%%%%%%%%%%%%%%%%%%%%%%%%%%%

\subsection{\label{DOS} Density of States}
To truly have complete information about a MMM system, we should know all of its possible energies. The number of states per interval of energy at each energy level that are available to be occupied is important for calculations of the effects based on the band theory like transition probability \cite{Harrison}. The actual transition probability depends on how many states are available in both the initial and final energies. The density of states (DOS) appears in calculations of conductivity where it presents the number of mobile states, and also in computing scattering rates where it presents the number of final states after scattering \cite{kuno,ashcroft}. DOS is a function of energy that when multiplied by an interval of energy, provides the concentration of the states which are available in that range. Before we can calculate the density of carriers in a polymer of interest, we have to find the number of available states at each energy. The Green's function of the hamiltonian of the MMM system is a key concept with important links to the concept of DOS via the following equation:

\begin{small}
\begin{align}
DOS(\epsilon) = \frac{-1}{\pi}\Im(Tr(G))
\end{align}
\end{small}

To calculate $ Tr(G) $, one can use Jacobi's formula\cite{Horn}, which expresses the derivative of the determinant of a matrix in terms of the adjugate of it. We hence conclude that

\begin{small}
\begin{align}
DOS(\epsilon) = \frac{-1}{\pi}\Im(\frac{d}{d\epsilon}ln \left| \widetilde{\mathcal{H}} \right|)
\end{align}
\end{small}

In which $ \left| \widetilde{\mathcal{H}} \right| $ would be calculated by the equation \ref{detH} which is a function of the monomer only.

%%%%%%%%%%%%%%%%%%%%%%%%%%%%%

\subsection{\label{LDOS} Local Density of States}
The work presented here will be given in terms of a physical quantity that describes the density of states, but space-resolved. Local variations, most often due to distortions of the original system, are often called local density of states (LDOS). We describe a method for calculation of the LDOS on polymers and finding a relation between LDOS of a desire polymer and that of its monomer. This relation can be used to gain keep insights into a subunit of the polymer as the main constituent. According to the formalism is used in the current work, one needs to calculate $ G_{i,i} $ in the following equation

\begin{small}
\begin{align}
LDOS_i(\epsilon) = \frac{-1}{\pi}\Im(G_{i,i})
\end{align}
\end{small}

To calculate $ G_{i,i} $, one needs to find the cofactor for that element. Let $ H^j $ stands for the hamiltonian of the monomer in which the values on that row and column are ignored. Subsequently, by the approach explained in the section \ref{method}, $ \widetilde{U}_j $ would be defined as follows

\begin{scriptsize}
\begin{align}
\widetilde{U}_j=\left(
\begin{array}{ccccccc}
 \left| H_{1:n-1}^j\right|  & -t_l{}^2\left| H_{1:n-2}^j\right|  & 0 & 0 & \cdots & 0 & 0 \\
 \left| H_{2:n-1}^j\right|  & -t_l{}^2\left| H_{2:n-2}^j\right|  & 0 & 0 & \cdots & 0 & 0 \\
 \vdots & \vdots & \vdots & \vdots & \cdots & \vdots & \vdots \\
 \left| H_{n-2:n-1}^j\right|  & -t_l{}^2\left| H_{n-2:n-2}^j\right|  & 0 & 0 & \cdots & 0 & 0 \\
 \left| H_{n-1:n-1}^j\right|  & -t_l{}^2 & 0 & 0 & \cdots & 0 & 0 \\
 1 & 0 & 0 & 0 & \cdots & 0 & 0 \\
\end{array}
\right)_{n\times n}
\end{align}
\end{scriptsize}

And also, for the Jth subunit one should obtain the following relation
\begin{scriptsize}
\begin{align}\label{HJj}
\left(
\begin{array}{c}
 \left| \mathcal{H}_1^{J,j}\right|  \\
 \left| \mathcal{H}_{2}^{J,j}\right|  \\
 \vdots \\
 \left| \mathcal{H}_{n-1}^{J,j}\right|  \\
 \left| \mathcal{H}_{n}^{J,j}\right|  \\
\end{array}
\right)=U^{J-1} \widetilde{U}_j U^{N-J-1}\left(
\begin{array}{c}
 \left| H_{1}\right|  \\
 \left| H_{2}\right|  \\
 \vdots \\
 \left| H_{n-1}\right|  \\
 \left| H_{n}\right|  \\
\end{array}
\right)
\end{align}
\end{scriptsize}
In which $ \left| \mathcal{H}_1^{J,j}\right| $ stands for the full hamiltonian in which the values on the jth row and jth column in the Jth subunit are ignored. After calculating the right hand side of the equation \ref{HJj} by $ J=1 $ one can obtain the LDOS for the first subunit of the monomer as follows

\begin{footnotesize}
\begin{align}
LDOS_j^{J=1}(\epsilon )=\frac{-1}{\pi }\Im\left(\frac{k_j\left(\lambda _1,\lambda _2\right)+k_j\left(\lambda _2,\lambda _1\right)}{\left|
\mathcal{H}\right| }\right)
\end{align}
\end{footnotesize}

In which $ k_j\left(\lambda _1,\lambda _2\right) $ is defined as following expression

\begin{widetext}
\begin{scriptsize}
\begin{multline}
k_j\left(\lambda _1,\lambda _2\right)=\frac{-\lambda _1{}^{N-2} }{4 \left(\lambda _1-\lambda _2\right) t_l^2 \left| H_2\right|
}\Big[\Sigma _r \left(\left| H\right| +\lambda _2-\lambda _1\right)+t_l^2 \left(\Sigma _r \left| H_{2:n-1}\right| +2 \left| H_2\right|
\right)\Big]\Bigg[ \left.2 \left| H_{1:n-1}\right|  \left(\Sigma _l \left| H_{2:n-1}^j\right| +\left| H_{1:n-1}^j\right|
\right)\right) -\\
\left| H\right|  \left(\left(\Sigma _l \left| H_{2:n-1}^j\right|  +\left| H_{1:n-1}^j\right| \right) \left(t_l^2 \left|
H_{2:n-1}\right| +\lambda _1-\lambda _2 + \left| H\right| \right) \right)-
 t_l^2 \left| H_2\right|  \left(\left(\Sigma _l \left| H_{2:n-2}^j\right|  +\left| H_{1:n-2}^j\right|  \right)\left(t_l^2
\left| H_{2:n-1}\right| -\lambda _1+\lambda _2 - \left| H\right| \right) \right. \left. \Bigg] \right.
\end{multline}
\end{scriptsize}
\end{widetext}

It should be noted that, presented expression is a function of the hamiltonian of the monomer $ H $. Just as a conclusion, LDOS for the Jth subunit can be calculated by the equation \ref{HJj} in terms of the first subunit as the form below

\begin{scriptsize}
\begin{align}
f_{J=1}\left(\lambda _1,\lambda _2\right) \left(\frac{\lambda _1}{\lambda _2}\right){}^{J-1}+f_{J=1}\left(\lambda _2,\lambda _1\right)
\left(\frac{\lambda _2}{\lambda _1}\right){}^{J-1}+c\left(\lambda _1,\lambda _2\right)
\end{align}
\end{scriptsize}

or more avail form

\begin{scriptsize}
\begin{align}
f_{J=1}\left(\lambda _1,\lambda _2\right) e^{i \omega  (J-1)}+f_{J=1}\left(\lambda _2,\lambda_1\right) e^{-i \omega (J-1)}+c\left(\lambda_1,\lambda _2\right)
\end{align}
\end{scriptsize}
In which $ \omega $ would be defined as follows

\begin{small}
\begin{align}
\omega =\tan ^{-1}\left(\frac{\Im\Delta}{\left| H_{2:n}\right|
-t_l^2 \left| H_{1:n-1}\right| }\right)
\end{align}
\end{small}

\begin{figure}[!htbp]
\centering
\includegraphics[width=0.9\linewidth]{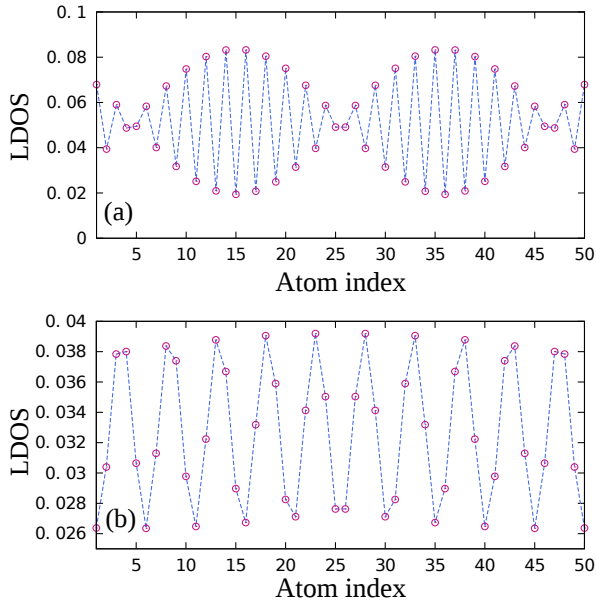}
\caption{\small{(a) and (b) Local density of states in different energies which have different frequencies.}}
\label{ff1}
\end{figure}

\begin{figure}[!htbp]
\centering
  \includegraphics[width=0.9\linewidth]{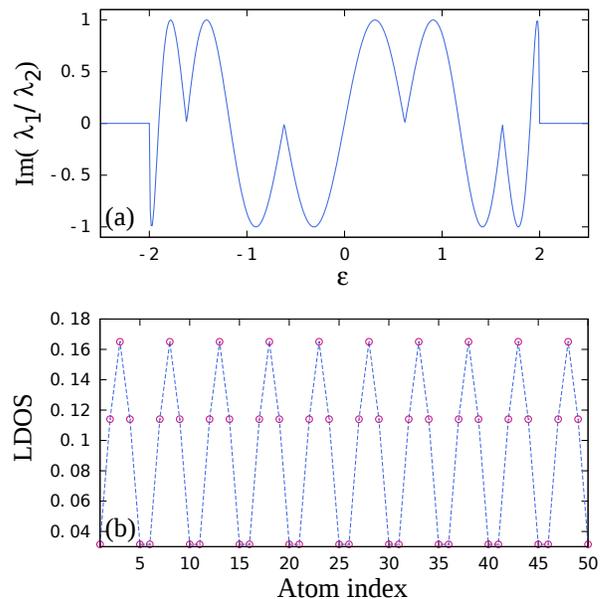}
\caption{\small{(a) The imaginary part of the fraction $ \lambda_1/\lambda_2 $, (b) Local density of states in a specific energy where the imaginary part of the fraction $ \lambda_1/\lambda_2 $ is zero.}}
\label{ff2}
\end{figure}

To clarify and show the importance of the issue, the variations of the LDOS of a typical polymer, contains five atoms in its monomer, is shown in figures \ref{ff1}a and \ref{ff1}b for different energies. As seen, $ \omega $ is the frequency of LDOS variations between subunits of the polymer. As shown in figure \ref{ff2}a, the imaginary part of the fraction $ \lambda_1/\lambda_2 $ in some specific energies is zero for which $ \omega $ would be equal to zero and LDOS should be uniform through the polymer. Figure \ref{ff2}b shows that, the LDOS (for a desire $ j $) for these energies would not be changed in pass from a subunit to the next one.

%%%%%%%%%%%%%%%%%%%%%%%%%%%%%%%%%%%%%%%%%%%%%%%%%%%%%%%%%%%%%%%%

\section{Summary}
\label{Summary}
In summary, a Metal-Molecule-Metal system has served as an interesting and widely used object of research in which the molecule is a polymer that is able to conduct charge. We have used the basic properties of matrices to calculate the determinant of the full hamiltonian of such a MMM system as a function of the determinant of the monomer. After that we have found transmission, density of states (DOS) and local density of states (LDOS) for the polymer as a function of the monomer and also, we have obtained a frequency for LDOS variations in pass from a subunit to the next one which is a function of energy. In some specific energies, that frequency is equal to zero and LDOS would be uniform through the polymer.

%\bibliography{mypaper}% Produces the bibliography via BibTeX.

\end{document}